\documentclass[12pt]{article}
\usepackage{amsmath,amssymb,mathrsfs,amsthm}
\usepackage{graphicx,bm,geometry,fullpage}
\usepackage{authordate1-4}

\newtheorem{theorem}{Theorem}

\newcommand{\abs}[1]{\left|#1\right|}
\newcommand{\mcl}[1]{\mathcal{#1}}

\newcommand{\startmat}[1]{\left(\begin{array}{#1}}
\newcommand{\closemat}{\end{array}\right)}

\newcommand{\bfP}{\mathbf{P}}
\newcommand{\bfM}{\mathbf{M}}
\newcommand{\defaultspread}{\linespread{1.5}}
\newcommand{\ffootnote}[1]{\linespread{1}\footnote{#1}\defaultspread}

\usepackage[font=small,width=\textwidth]{caption}
\title{Two-Stage Bayesian Model Averaging in Endogenous Variable Models\footnote{We thank David Albouy and Francesco Trebbi for kindly sharing their data as well as Chris Papageorgiou for helpful comments. Lenkoski's research is supported by DFG research grant 1653. Raftery's research
was supported by NIH Grants ~ R01 HD-54511 and R01~GM-084163, NSF grant no.~llS-0534094, 
and NSF grant no.~ATM-0724721.}}
\date{}
\author{\textsc{By Alex Lenkoski}\\ {\it Institute for Applied Mathematics, Heidelberg University} \\\textsc{Theo S. Eicher}\\ {\it Department of Economics, University of Washington}\\\textsc{and Adrian E. Raftery}\\ {\it Department of Statistics, University of Washington}\\
}

\begin{document}

\maketitle
\begin{abstract}
Economic modeling in the presence of endogeneity is subject to model uncertainty at both the instrument and covariate level. We propose a Two-Stage Bayesian Model Averaging (2SBMA) methodology that extends the Two-Stage Least Squares (2SLS) estimator. By constructing a Two-Stage Unit Information Prior in the endogenous variable model, we are able to efficiently combine established methods for addressing model uncertainty in regression models with the classic technique of 2SLS.  To assess the validity of instruments in the 2SBMA context, we develop Bayesian tests of the identification restriction that are based on model averaged posterior predictive $p$-values. 
A simulation study showed that 2SBMA has the ability to recover structure in both the instrument and covariate set, and substantially improves the sharpness of resulting coefficient estimates in comparison to 2SLS using the full specification in an automatic fashion.  Due to the increased parsimony of the 2SBMA estimate, the Bayesian Sargan test had a power of 50 percent in detecting a violation of the exogeneity assumption, while the method based on 2SLS using the full specification had negligible power. We apply our approach to the problem 
of development accounting, and find support not only for institutions, but also for geography and integration as development determinants, once both model uncertainty and endogeneity have been jointly addressed.  
\end{abstract}

\newpage
\baselineskip=18pt


\section{Introduction}
\indent In modeling economic interactions, applied economists are often faced with a surfeit of theories and a number of different variables that proxy a given theory. Raftery \shortcite{raftery_1995} argued that uncertainty surrounding particular theories should be addressed explicitly by the statistical approach. 
Uncertainty about a parameter can be underestimated when it is based on a single
model and there is uncertainty about which model or theory to use.
Bayesian model averaging (BMA) has been used extensively to account for model uncertainty in regression models, including in cross-country growth regressions and development accounting.\ffootnote{See e.g., Fernandez et al. \shortcite{fernandez_et_2001}, Sala-i-Martin et al. \shortcite{sala-i-martin_et_2004}, Ciccone and Jarocinski \shortcite{ciccone_jarocinski_2007}, and Eicher et al. \shortcite{eicher_et_2007}.}  However, these methods are constrained to modeling assocations between covariates and the dependent variable, ignoring issues of endogeneity and uncertainty surrounding instrument specifications.\\
\indent We propose a new methodology that combines the BMA approach to regression variable uncertainty and the Two-Stage Least Squares (2SLS) procedure.  Our new method, Two-Stage BMA (2SBMA), enables the applied economist to address uncertainty in instrument and covariate specifications in a natural manner while simultaneously yielding an estimator that addresses endogeneity in the covariates.  We propose this technique as an automatic alternative to the standard 2SLS exercises, which often involve extensive, and typically unreported remodeling in order to obtain a reduced, relevant, covariate set.\\
\indent To date, instrument uncertainty has been addressed only in standard robustness analyses that juxtapose instruments associated with one particular theory/specification against another. In a prominent example in the development accounting literature, Rodrik et al. \shortcite{rodrik_et_2004}, henceforth RST, motivate their work by a ``horse race" among alternative theories that propose candidate instruments and regressors.\\
\indent Accounting for uncertainty about both covariates and
instruments requires a methodology that is rooted in statistical theory. Durlauf et al. \shortcite{durlauf_et_2007} introduced an instrument model selection procedure based on evaluating coefficient estimates according to $t$-statistics, but warned of the tenuous nature of the underlying statistical theory. 
The most comprehensive approach to addressing endogeneity in growth regressions has previously been proposed by Durlauf et al. \shortcite{durlauf_et_2008}, who built on Tsangarides \shortcite{tsangarides_2004}. The authors introduced a model averaged version of 2SLS, but noted that their heuristic approach lacked statistical justification.\ffootnote{A similar heuristic panel approach is introduced by Hineline \shortcite{hineline_2007} to examine the growth/inflation relationship. 
Morales-Benito (2009) provides statistical foundations for a panel BMA approach.}  Strictly speaking, Durlauf et al. \shortcite{durlauf_et_2008} did not allow for instrument uncertainty, but provided a model averaging approach to instrument candidate regressors in the second stage only.\\
\indent The 2SBMA approach extends the method of Durlauf et al. \shortcite{durlauf_et_2008} and provides a statistical foundation for the procedure. An alternative to theory-specified instruments is the use of lagged dependent variables in a BMA-GMM approach (see Morales-Benito \shortcite{morales-benito_2009}).  We explore the properties of 2SBMA as a valid two-stage estimator and show that the procedure is a consistent methodology that reduces the well known many-instrument bias present in 2SLS. A simulation study reported below shows a $45\%$ reduction in many instrument bias, as well as a 46\% reduction in mean squared error for estimating regression coefficients as opposed to 2SLS using the full specification.\\
\indent Instrumental variable estimation requires assumptions that relate to the identification of the implied structural model.  
Tests of the identification restrictions, such the Sargan \shortcite{sargan_1958} and Cragg and Donald \shortcite{cragg_donald_1993} tests, compare a test statistic to a  reference distribution.  The test statistics are often asymptotically distributed according to the reference distribution, which frequently has a number of degrees of freedom that is related to the size of the model estimated.  The nature of these statistics proves problematic in some economic applications, such as the study of growth and development determinants, where the  sample size is small and additional regressors continue to be proposed.\\
\indent The 2SBMA approach provides a direct interpretation of the efficacy of an instrumentation strategy, by examining posterior inclusion probabilities. However, we also provide measures to verify instrument conditions based on Bayesian model averaged posterior predictive $p$-values.  
We introduce these $p$-values as a simple extension of the work of Rubin \shortcite{rubin_1984}, Raftery \shortcite{Raftery1988} and Gelman et al. \shortcite{gelman_et_1996}, who motivated the use of posterior predictive $p$-values for a single model. These $p$-values are then used to derive Bayesian versions of the 
Sargan and Cragg and Donald tests to test the identification restrictions. \\
\indent A simulation study quantifies the efficiency of 2SBMA as compared to 2SLS. We found that the Bayesian over-identification test had a power of 50\% at detecting this failure, while the traditional Sargan test had a power of only 0.8\%.  We note that this results from the increased parsimony that the 2SBMA estimator obtains in comparison  to the 2SLS estimator, highlighting the fact that 2SBMA is able to automatically introduce beneficial levels of parsimony in the modeling exercise.\\
\indent Finally, we apply 2SBMA to a prominent approach to development accounting that is known to be subject to substantial instrument and determinant uncertainty.\ffootnote{We use the term development accounting in the broad sense, referring to studies that seek to examine the determinants of differences in levels of per capita income. Previous development accounting approaches differ in their emphases, such as physical capital (King and Levine \shortcite{king_levine_1994}), human capital (Klenow et al. \shortcite{klenow_et_1997}), as well as TFP (Caselli \shortcite{caselli_2005}). In our application we focus only on studies that sought to explain to differences in per capita incomes based on integration, institutions, and geography since the above approaches do not feature instrument uncertainty.} 
We use RST's own data and the variables motivated by their empirical approach to span the model space for our 2SBMA approach. RST found strong evidence for the ``primacy of institutions'' over all other alternative theories after conducting an elaborate ``horse race" among all alternative theories and their associated  candidate regressors. The 2SBMA results suggest a qualification of RST's strong conclusions. We find that not only institutions, but also integration and geography explain differences in per capita incomes across countries. This suggests that previously reported robustness results may have depended largely on which specifications were used. In addition we find that a number of instruments that are associated with alternative theories also receive support from the data, once instrument and determinant uncertainty have been addressed simultaneously.
These results are consistent with the findings of Durlauf et al. \shortcite{durlauf_et_2008} that many growth theories/variables are not robust once model uncertainty is taken into account. \\
\indent Through his work on Bayesian regression and time series, Arnold Zellner had considerable influence on the material presented here.  In particular, one of the main themes of this work is proposing a proper prior distribution on the regression parameters.  The g-prior of Zellner \shortcite{zellner_1986} is often used for model selection exercises in the regression context because it is a proper prior distribution.  In the model we consider below, we extend the Unit Information Prior \cite{kass_wasserman_1995}, which can be seen as a specific version of Zellner's g-prior. Arnold Zellner was also an early contributor to the literature on Bayesian combination of forecasts and models \cite{palm_zellner_1992,min_zellner_1993}, which underlies much of the present work.\\
\indent The article proceeds as follows. Section 2 outlines the statistical foundations of the 2SBMA approach. Section 3 introduces Bayesian tests of the identification restrictions and Section 4 describes the efficiency of 2SBMA and the power of Bayesian tests in simulation studies. Section 5 applies 2SBMA to a prominent approach to development accounting to highlight the importance of both determinant and instrument uncertainty.  Section 6 concludes.\\
\section{Statistical Foundations of 2SBMA}
This section develops the theoretical foundations for two-stage Bayesian Model Averaging (2SBMA). First, we sketch the properties of the 2SLS estimator that we extend.  Then we formulate the endogeneity problem to express first and second stage likelihoods in a model averaging context. After developing a two-stage prior that allows for an efficient approximation of the model probabilities, we then derive the complete 2SBMA methodology.\\ 
\subsection{Marginal and Conditional Likelihoods for Two-Stage Applications}
A standard approach to addressing endogeneity is to apply 2SLS and impose identification restrictions.  We consider the model,
\begin{align}
\label{eq:model1}
Y &= (W\mbox{ }X) \beta+ \eta , \\
\label{eq:model2}
W &= (Z\mbox{ }X)\theta + \epsilon ,
\end{align}
where $Y$ is the $n \times 1$ dependent variable, $X$ is an $n \times $ $p_X$ set of covariates, $W$ is the set of endogenous variables and $Z$ is the $n \times p_Z$ set of instruments.  The notation $(A\mbox{ }B)$ denotes a matrix placing A and B together, provided they have the same number of rows. To simplify exposition we assume that 
$W$ is $n \times 1$. Assuming that
\[
\startmat{c}\eta\\\epsilon\closemat \sim N\left(\startmat{c}0\\0\closemat,\startmat{c c}\sigma^2_{\eta} & \sigma_{\eta\epsilon}\\\sigma_{\eta\epsilon} & \sigma^2_{\epsilon}\closemat = \Sigma\right),
\]
the classical endogenous variable situation arises when $\sigma_{\eta\epsilon}\neq 0$, causing $W$ to violate the standard regression assumption of independence of the error term, $\eta$.\\ 
\indent Fundamental to the 2SBMA methodology is the specification of two-stage marginal and conditional likelihoods. Letting $U = (W\mbox{ }X)$ and $V = (Z\mbox{ }X)$, we adopt the notation of Kleibergen and Zivot \shortcite{kleibergen_zivot_2003}, and rewrite Equations~\ref{eq:model1} and \ref{eq:model2} as
\renewcommand{\theequation}{\arabic{equation}'} 
\setcounter{equation}{0}  
\begin{align}
\label{eq:rry}
Y &= \tilde{U}(\theta)\beta + \nu, \\
\label{eq:rrw}
W &= V\theta + \epsilon,
\end{align}
where $\tilde{U}(\theta) = (V\theta\mbox{ }X)$ is the replacement of $W$ in $U$ with its fitted value, and the OLS estimate regressing W on V is given by   $\hat{U} = \tilde{U}(\hat{\theta})$ where $\hat{\theta} = (V'V)^{-1}V'W$. Since $\nu = \beta\epsilon + \eta$, we have 
\[
\text{Var}\startmat{c}\nu\\\epsilon\closemat = \Omega = \startmat{cc}\omega_{11} & \omega_{12}\\\omega_{21} & \omega_{22}\closemat = \startmat{cc}1 & 0\\\beta & 1\closemat'\Sigma\startmat{cc}1 & 0\\\beta & 1\closemat.
\]
Kleibergen and Zivot \shortcite{kleibergen_zivot_2003} reparameterize (\ref{eq:rry}) and (\ref{eq:rrw}) using $\phi = \omega_{22}^{-1}\omega_{21}$ to yield
\renewcommand{\theequation}{\arabic{equation}'{'}} 
\setcounter{equation}{0}  
\begin{align}
\label{eq:ykz}
Y &= \tilde{U}(\theta)\beta + \xi + \epsilon\phi, \\
\label{eq:xkz}
W &= V\theta + \epsilon,
\end{align}
\renewcommand{\theequation}{\arabic{equation}} 
\setcounter{equation}{2}  
where $var(\xi) = \omega_{11\cdot 2} = \omega_{11} - \omega_{12}^2\omega_{22}$. \\
\indent In the presence of endogeneity, it is well known that the correlation between $\nu$ and $\epsilon$ leads to inconsistent estimates of the entire coefficient vector, $\beta$, under standard Ordinary Least Squares (OLS). The 2SLS estimator, 
\[
\hat{\beta}^{2SLS} = \left(\hat{U}'\hat{U}\right)^{-1}\hat{U}'Y,
\]
may resolve the inconsistency problem when instruments, $Z$, exist that are independent of $Y$, given $W$ and the vector of covariates, $X$. The model is identified only when these conditional independence assumptions are valid, resulting in a consistent estimator. The 2SLS estimate of $\beta$ is well known to be biased, and the extent of this bias increases with the number of terms that are added in the first stage with coefficients equal, or close to, zero. Under such conditions, 2SLS estimates  may not reduce the bias of the OLS results \cite{davidson_mackinnon_2004}.  It is therefore of practical interest to develop a 2SBMA methodology that quantifies the degree to which each proposed covariate and instrument enters into the equations above.\\
\indent The reparameterization of (\ref{eq:ykz}) and (\ref{eq:xkz}) proves useful to derive the two-stage marginal and conditional likelihoods. 
\begin{theorem}\label{thm:MLE} The two-stage marginal and conditional likelihoods are given by 
\begin{align}
\label{eq:marginalW}
L(\theta|W, V, \omega_{22}) &= \omega_{22}^{-n/2}\exp\left(-\frac{1}{2}\frac{(W - V\theta)'(W - V\theta)}{\omega_{22}}\right)\\
\label{eq:condX}
L(\beta|\phi, \omega_{11\cdot 2}, Y, \hat{\theta}) &= \omega_{11\cdot 2}^{-n/2}\exp\left(-\frac{1}{2} \frac{(Y - \hat{U}\beta + \hat{\epsilon}\phi)'(Y - \hat{U}\beta + \hat{\epsilon}\phi)}{\omega_{11\cdot 2}}\right),
\end{align}
where $\hat{\epsilon} = W - V\hat{\theta}$. The marginal likelihood is maximized at $\hat{\theta} = (V'V)^{-1}V'W$, while the conditional likelihood is maximized at $\hat{\beta}^{2SLS}$.
\end{theorem}
\emph{Proof} See Appendix. \\
\subsection{Constrained Two-Stage Likelihoods}
\indent When juxtaposing competing theories using 2SBMA, some coefficients in either $\theta$ or $\beta$ are constrained to be zero. As a consequence, these members of either $X$ or $Z$ need not be included in estimation of the associated parameter. This section derives the constrained marginal and conditional likelihoods for this case.\\
\indent Let $\mcl{M}$ be a collection of individual first stage models,
$\{M_1, \dots, M_I\}$.  Associated with each first stage model $M_i$ is a submatrix $V^{(i)}$ of the matrix $V$.  When the variables $Z_l$ and $X_m$ are in $M_i$, then these columns are retained in $V^{(i)}$, while those not in $M_i$ are excluded. For $M_i$, the least squares estimate is $\hat{\theta}^{(i)} = \left(V^{(i)'}V^{(i)}\right)^{-1}V^{(i)'}W$.\\
\indent Similarly, let $\mcl{L}$ be a set of individual second stage models, $\{L_1, \dots, L_J\}$.  Each model $L_j$ again  corresponds to a submatrix $U^{(j)}$ of the matrix $U$.  For $W\in L_j$, $\tilde{U}^{(j)}(\theta^{(i)})$ denotes the replacement of $W$ in $U^{(j)}$ with the fitted value $V^{(i)}\theta^{(i)}$, and when $W\notin L_j$, we write  $\tilde{U}^{(j)}(\theta^{(i)}) = U^{(j)}$. To ease notation, we write $\hat{U}^{(i,j)}$ for $\tilde{U}^{(j)}(\hat{\theta}^{(i)})$.\\
\indent Analogous to the representation in Equations~\ref{eq:ykz} and \ref{eq:xkz}, the sampling model under $M_i$ and $L_j$ takes the form
\begin{align}
\label{eq:ykz_constr}
Y &= \tilde{U}^{(j)}(\theta^{(i)})\beta^{(j)} + \xi^{(i,j)} + \epsilon^{(i)}\phi^{(i,j)}, \\
\label{eq:xkz_constr}
W &=  V^{(i)}\theta^{(i)} + \epsilon^{(i)},
\end{align}
where $var(\epsilon^{(i)}) = \omega_{22}^{(i)}$, $var(\xi^{(i,j)}) = \omega_{11\cdot 2}^{(i,j)}$. The constrained 2SLS estimator based on models $M_i$ and $L_j$ is then calculated as
\begin{equation}\label{eq:const2sls}
\hat{\beta}^{(i,j)} = \left(\hat{U}^{(i,j)'}\hat{U}^{(i,j)}\right)^{-1}\hat{U}^{(i,j)'}Y .
\end{equation}
\indent Theorem 2 highlights how the marginal and conditional likelihoods are affected when a variable contained in $U^{(j)}$ is excluded from from $V^{(i)}$.
\begin{theorem}\label{thm:MLEConstr} Under models $M_i$ and $L_j$ the expressions in Equations~(\ref{eq:ykz_constr}) and (\ref{eq:xkz_constr}) 
yield the following marginal and conditional likelihoods:
\begin{align}
\label{eq:marginalW}
L\left(\theta|W, V, \omega^{(i)}_{22}\right) &= \left(\omega^{(i)}_{22}\right)^{-n/2}\exp\left(-\frac{1}{2}\frac{(W - V^{(i)}\theta^{(i)})'(W - V^{(i)}\theta^{(i)})}{\omega^{(i)}_{22}}\right) , \\
\label{eq:condX}
L\left(\beta|\phi^{(i,j)}, \omega^{(i,j)}_{11\cdot 2}, Y, \hat{\theta}^{(i)}\right) &= \left(\omega^{(i,j)}_{11\cdot 2}\right)^{-n/2}\exp\left(-\frac{1}{2} \frac{(Y - \hat{U}^{(j)}\beta^{(j)} + \hat{\epsilon}^{(i)}\phi^{(i,j)})'(Y - \hat{U}^{(j)}\beta^{(j)} + \hat{\epsilon}^{(i)}\phi^{(i,j)})}{\omega^{(i,j)}_{11\cdot 2}}\right),
\end{align}
where $\hat{\epsilon}^{(i)} = W - V^{(i)}\hat{\theta}^{(i)}$.  Equation~(\ref{eq:marginalW}) is maximized at $\hat{\theta}^{(i)}$, while Equation~(\ref{eq:condX}) is maximized at
\begin{equation}\label{eq:trueMLE}
\hat{\beta}^{(i,j)} + \phi^{(i,j)}\hat{\Pi}^{(i,j)},
\end{equation}
where $\hat{\Pi}^{(i,j)} = (\hat{U}^{(j)'}\hat{U}^{(j)})^{-1}\hat{U}^{(j)'}\hat{\epsilon}^{(i)}$.
\end{theorem}
\emph{Proof} See appendix.\\
\indent Theorem~\ref{thm:MLEConstr} implies that the mode of the conditional likelihood for $\beta$ equals $\hat{\beta}^{(i,j)}$ plus the term $\phi^{(i,j)}\hat{\Pi}^{(i,j)}$. To interpret this difference we note that $\hat{\Pi}^{(i,j)}$ represents the regression coefficients of $\hat{U}^{(j)}$ on the first stage residuals. Hence for the case where all covariates in $L_j$ are contained in $M_i$, $\hat{\Pi}^{(i,j)} = 0$ and Equation (\ref{eq:trueMLE}) becomes  $\hat{\beta}^{(i,j)}$. For the case where variable $X_l$ is in $L_j$ but excluded from $M_i$, and where $X_l$ has little explanatory power for $W$ given the remaining elements of $M_i$, $\hat{\Pi}^{(i,j)}$ is, by definition, negligible.  The 2SBMA assumptions outlined below that govern the model pair $(M_i, L_j)$ constrain $\hat{\Pi}^{(i,j)}$ to zero.\\
\indent The 2SBMA methodology is based on calculating the quantities in Equation~(\ref{eq:const2sls}). As opposed to Limited Information Maximum Likelihood (LIML), the focus of our estimator is on maximizing the marginal, then the conditional likelihoods, yielding estimators of the form $\hat{\beta}^{(i,j)}$, while LIML maximizes the joint likelihood of the parameters.  This causes our methodology to be more similar to 2SLS than LIML.  Indeed, note that when $M_i$ and $L_j$ are the full models for the first and second stage, respectively, $\hat{\beta}^{(i,j)} = \hat{\beta}^{2SLS}$. Furthermore, if $X_l$ is the only variable excluded from $M_i$ and $L_j$, $\hat{\beta}^{(i,j)}$ is simply the 2SLS estimate with this covariate excluded.  Similarly, if $Z_m$ is the only quantity excluded from $M_i$, and $L_j$ is the full model, $\hat{\beta}^{(i,j)}$ is the 2SLS estimate with instrument $Z_m$ excluded.\\
\indent The formulation above also allows for two occurrences not considered in standard 2SLS.  First, we may have variable $X_l\in M_i$, but $X_l\notin L_j$.  This is equivalent to stating that a variable originally considered to be a covariate is now considered conditionally independent of $Y$, given the remaining variables in $L_j$. However, the variable still has some explanatory power for $W$, which causes it to be included in $M_i$. Second, we may include variable $X_l \in L_j$, but exclude it from  $M_i$.  This case is not considered in standard 2SLS analysis, as it is well known to result in bias if the effect of $X_l$ on $W$ is substantial, as outlined by the term $\hat{\Pi}^{(i,j)}$ in Theorem~\ref{thm:MLEConstr} above.  In 2SBMA we do allow for these occurrences, since a large $\hat{\Pi}^{(i,j)}$ would imply a failure in the model selection methodology during the first stage.
\subsection{The Two-Stage Unit Information Prior}
\indent 2SBMA requires a two-stage prior distribution over the parameters of the model.  In the context of Bayesian Instrumental Variable estimation, Dreze \shortcite{dreze_1976} suggested the improper prior 
\[
pr(\beta, \theta, \Sigma) \propto \abs{\Sigma}^{-1/2},
\]
for Equations (\ref{eq:model1}) and (\ref{eq:model2}). This two-stage prior is in part motivated by the standard improper prior $pr(\beta, \sigma) \propto 1/\sigma$ placed on Bayesian regression problems.\\
\indent Kleibergen and Zivot \shortcite{kleibergen_zivot_2003} note that the Dreze prior may be poorly behaved as $\theta \to 0$. Instead they suggest 
\begin{equation}\label{eq:kzprior}
pr(\beta, \theta, \phi, \omega_{11\cdot 2}, \omega_{22}) \propto \omega_{11\cdot 2}^{-1} \abs{\Omega}^{-1/2}\abs{\theta'V'V\theta}^{1/2},
\end{equation}
a prior that includes dependence on the first stage coefficients, $\theta$. Kleibergen and Zivot \shortcite{kleibergen_zivot_2003} note that the posterior from this prior bears a number of similarities to the 2SLS estimator.  In particular, the conditional posterior of $\beta$ when $\theta = \hat{\theta}$ has mean and mode equal to $\hat{\beta}^{2SLS}$.\\
\indent While the Kleibergen and Zivot prior has a number of desirable properties, it is still improper in several parameters.  This may not have undue influence on standard Bayesian posterior parameter estimation. Kass and Raftery \shortcite{kass_raftery_1995} note, however, that the use of improper priors has a greater impact in the context of model comparison. Hence we seek to develop a  two-stage prior which shares many of the properties of the Kleibergen and Zivot \shortcite{kleibergen_zivot_2003} prior, but is proper.  We choose to base our analysis on the Unit Information Prior (UIP), which has been motivated in the context of BMA \cite{kass_wasserman_1995,raftery_1995}. 
The UIP is a normal prior with a mean centered at the maximum likelihood estimate and a variance equal to the inverse of the average information contained in one observation.  \\
\indent We construct a prior of the form
\[
pr(\beta, \theta, \omega_{11\cdot2}, \omega_{22}, \phi) = pr(\beta, \omega_{11\cdot 2}, \phi|\theta, \omega_{22})pr(\theta,\omega_{22}),
\]
and specify a standard UIP on $pr(\theta, \omega_{22})$, based on the likelihood in Equation (\ref{eq:marginalW}).  In the second stage, we construct a prior that is conditional on the first stage, 
$pr(\beta, \omega_{11\cdot 2}, \phi|\theta, \omega_{22})$, 
and specify a UIP for given values of $(\theta,\omega_{22})$, based on the likelihood in Equation~(\ref{eq:condX}).  By Theorem~\ref{thm:MLE}, the resulting two-stage UIP centers $\beta$ at  $\hat{\beta}^{2SLS}$ when $\theta$ is set to its mode, $\hat{\theta}$.\\
\indent A prior over parameters of restricted models, $(M_i, L_j)$, is specified in essentially the same manner
\[
pr\left(\beta^{(j)}, \theta^{(i)}, \omega^{(i,j)}_{11\cdot2}, \omega^{(i)}_{22}, \phi^{(i,j)}\right) = pr\left(\beta^{(j)}, \omega^{(i,j)}_{11\cdot 2}, \phi^{(i,j)}|\theta^{(i)}, \omega^{(i)}_{22}\right)pr\left(\theta^{(i)},\omega^{(i)}_{22}\right),
\]
where a UIP is again specified on $pr(\theta^{(i)},\omega^{(i)}_{22})$.  Then, conditional on a value of $\theta^{(i)}$, we specify a UIP on $\beta^{(j)}$.\\
\indent By imposing the conditional independence assumption on the pair of models $(M_i, L_j)$, we can set $\hat{\Pi}^{(i,j)} = 0$.  Therefore, when $\theta^{(i)} = \hat{\theta}^{(i)}$, the prior distribution for $\beta^{(j)}$ is centered about $\hat{\beta}^{(i,j)}$.  This implies that the two-stage UIP can be considered a proper analogue of the Kleibergen and Zivot \shortcite{kleibergen_zivot_2003} prior. That is, the prior features a built-in dependence on $\theta$, while its posterior mimics the 2SLS estimator.
\subsection{Statistical Foundations of 2SBMA}\label{sec:2SBMA}
\indent 2SBMA combines the 2SLS methodology discussed above with the standard BMA methodology reviewed below. 2SBMA processes the data much like a two-stage estimator while addressing model uncertainty in both stages. The first stage is a simple application of BMA to identify effective instruments. It is helpful to review the properties of BMA that are implied in stage 1.\\
\indent Let $\Delta$ be a quantity of interest. In BMA, the posterior distribution of $\Delta$  given the data, D, is given by the weighted average of the predictive distributions under each model, weighted by the corresponding posterior probabilities,
\[
pr(\Delta|D) = \sum_{i  = 1}^{I} pr(\Delta|M_i, D) pr(M_i|D),
\]
\noindent where $pr(\Delta|M_i, D)$ is the predictive distribution given model $M_i$ and $pr(M_i|D)$ is the posterior model probability of model $M_i$.  The posterior model probability $\pi_i$, for each first stage model $M_i$ is given by
\begin{eqnarray*}
\pi_i & = &  pr(M_i|D) \\
& \propto &  pr(D|M_i)pr(M_i) ,
\end{eqnarray*}
where
\[
pr(D|M_i) = \int pr(D|\theta^{(i)}, M_i) pr(\theta^{(i)}| M_i) d\theta^{(i)}
\]
is the integrated likelihood of model $M_i$ with parameters $\theta^{(i)}$.  The prior densities for parameters and models are $pr(\theta^{(i)}|M_i)$ and $pr(M_i)$, respectively.\\
\indent Under BMA, the posterior mean of $\theta$ is the sum of the posterior means of each model in the collection $\mcl{M}$, weighted by their posterior probabilities
\[
\hat{\theta}^{BMA} = \sum_{i = 1}^{I} \pi_i\hat{\theta}^{(i)}.
\]
Similarly, the posterior variance is
\[
\sum_{i = 1}^{I}\pi_i\hat{\sigma}_i^2 + \sum_{i = 1}^{I}\pi_i\left(\hat{\theta}^{(i)} - \hat{\theta}^{BMA}\right)^2.
\]
The posterior variance highlights how BMA methodology accounts for model uncertainty. The first term is the weighted variance for each model, $\hat{\sigma}^2_i = Var(\hat{\theta}^{(i)}|M_i, D)$, averaged over all relevant models, and the second term indicates how stable the estimates are across models. The more the estimates differ between models, 
the greater is the posterior variance.\\
\indent The posterior distribution for a parameter is a mixture of a regular posterior distribution and a point mass at zero, which represents the probability that the parameter equals zero. The sum of the posterior probabilities of the models that contain the variable is called the inclusion probability.  For instance, for instrument $Z_{k}$ we may write,
\[
\mu^{BMA}(\theta_{Z_k}) = pr(\hat{\theta}_{Z_k}\neq 0|D) = \sum_{i \in \mcl{M}_k} \pi_i,
\]
where $\mcl{M}_{k}$ is a collection of indices for which $i \in \mcl{M}_k$ implies model $M_i$ does not restrict the parameter $\theta_{Z_k}$ to zero. 
Standard rules of thumb for interpreting $\mu^{BMA}$  have been provided by 
Kass and Raftery \shortcite{kass_raftery_1995}, modifying an earlier
proposal of Jeffreys \shortcite{jeffreys_1961}.
They suggested the following effect thresholds: 
$<$50\%: evidence against the effect, 50-75\%: weak evidence for the effect, 
75-95\%: positive evidence, 95-99\%: strong evidence, and $>$99\%: very strong evidence\ffootnote{From a decision theory perspective, one can imagine cases where a policy-maker might be interested in a variable even when its posterior inclusion probability is below 50\% ; see Brock and Durlauf \shortcite{brock_durlauf_2001} and Brock et al. \shortcite{brock_et_2003}}.\\
\indent In the case of 2SLS estimation in the presence of model uncertainty, the BMA framework must be extended to account for model uncertainty at both stages.  For models $M_i$ and $L_j$, we are now interested in
\begin{equation}\label{eq:modelsep}
pr(M_i, L_j|D) = pr(L_j|M_i, D)pr(M_i|D).
\end{equation}
The decomposition in Equation~(\ref{eq:modelsep}) indicates the dependence that the probability of model $L_j$ has on the particular choice of model $M_i$, as each model $M_i$ yields a slightly different distribution for the fitted value of $W$.  Furthermore, for a given value of $\theta^{(i)}$, the quantity $pr(L_j|\theta^{(i)}, M_i, D)$ is itself a regression problem with a particular UIP.\\
\indent Denoting by $\nu_{i,j}$ the probability of model $L_j$ given model $M_i$ yields the 2SBMA estimator
\begin{equation}\label{eq:2SBMA_estimate}
\hat{\beta}^{2SBMA} = \sum_{i = 1}^{I}\sum_{j = 1}^{J} \pi_i \nu_{i,j} \hat{\beta}^{(i,j)}.
\end{equation}
Equation~(\ref{eq:2SBMA_estimate}) shows that the 2SBMA estimate is formed as the average of each constrained 2SLS estimate that results from the combination of model $M_i$ in the first stage, and model $L_j$ in the second stage, weighted by both the first and second stage model probabilities. To calculate the posterior variance of $\hat{\beta}^{2SBMA}$ we have the following result.
\begin{theorem}\label{thm:2SBMA_var} Let $\hat{\beta}^{(i,\cdot)} = \sum_{j = 1}^{J} \nu_{i,j}\hat{\beta}^{(i,j)}$ be the model averaged estimate of $\beta$ for a given first stage model $M_i$. Then the variance of the estimate $\hat{\beta}^{2SBMA}$ is
\begin{equation}\label{eq:2SBMA_var_alt}
\sigma^2_{2SBMA}(\beta) = \sum_{i = 1}^{I} \pi_i Var(\beta|M_i) + \sum_{i = 1}^{I} \pi_i \left(\hat{\beta}^{(i,\cdot)} - \hat{\beta}^{2SBMA}\right)^2,
\end{equation}
where
\[
Var(\beta|M_i) = \sum_{j = 1}^{J} \nu_{i,j}Var({\beta}|M_i,L_j) + \sum_{j = 1}^{J} \nu_{i,j}\left(\hat{\beta}^{(i,j)} - \hat{\beta}^{(i,\cdot)}\right)^2,
\]
is the BMA variance associated with second stage estimates for a fixed first stage model.
\end{theorem}
\emph{Proof} See Appendix. \\
\indent Theorem~\ref{thm:2SBMA_var} shows that the variance of 2SBMA estimates can be decomposed into two parts that yield interpretations similar to standard BMA variances. The first term is the average of BMA variances associated with the models in the first stage, and the second term represents the variation of a given first stage model's BMA estimates relative to the overall 2SBMA estimate.\\
\indent The posterior distribution of $\hat{\beta}^{2SBMA}$ is again a mixture of a regular posterior distribution and a point mass at zero, which represents the probability that the parameter equals zero. The sum of these posterior probabilities that contain the variable is then the inclusion probability at the second stage.  For instance, for the variable $X_l$ we have
\[
\mu^{2SBMA}(\beta_{X_l}) = pr(\hat{\beta}^{2SBMA}_{X_l}\neq 0|D) = \sum_{i = 1}^{I}\sum_{j \in \mcl{L}_l} \pi_i \nu_{i,j}, 
\]
where $\mcl{L}_{l}$ is the subset of $\mcl{L}$ for which the coefficient  $\beta_{X_l}$ is not constrained to zero.  We continue to follow the standard rules of thumb for interpreting effect thresholds in the second stage[AR1]. \\
\indent A desirable feature of the 2SBMA estimator is that asymptotically it will resemble the 2SLS estimator.  This is due to the fact that any proposed instrument or covariate that has a zero coefficient in either the first of second stage will be dropped in the limit by the model selection procedure.  This yields the following result.
\begin{theorem} \label{thm:consistency}
The 2SBMA estimate is consistent under identification in the full model.  $\hat{\beta}^{2SBMA} \to_p \beta$ when $\hat{\beta}^{2SLS} \to_p \beta$.
\end{theorem}
\emph{Proof} See Appendix.
\noindent This result mainly implies that one does not sacrifice consistency when employing 2SBMA rather than 2SLS.\\
\indent We further note that coefficients derived from instrumental variables
regressions exhibit increasing finite sample bias as the number of instruments increases (see, e.g., Hall \shortcite{hall_2005}). The bias is aggravated when the proposed instruments have little explanatory power for the endogenous variable (Bound et al. \shortcite{bound_et_1995}). While it is difficult to assess in general, the 2SBMA methodology has the potential to limit this finite sample bias, by dropping unnecessary terms.  The simulation study in Section 4 shows an instance in which this occurs.\\

\section{Bayesian Tests of Identification Restrictions}

\indent Various tests have been developed to examine the instruments' conditional independence (on $Y$) and explanatory power (for $W$).  Below we develop Bayesian equivalents of the Sargan \shortcite{sargan_1958} test of over-identification and the Cragg and Donald  \shortcite{cragg_donald_1993} test of  under-identification that Stock and Yogo \shortcite{stock_yogo_2002} used to propose a weak instruments test.  We focus on these tests as they are widely in use by the applied community and have proven robust at detecting violations of the identification assumptions.  Further, as shown in the simulation study of Section~\ref{sec:Sim}, they perform well in our context. We then show how model averaged versions of these tests can be used in the 2SBMA framework to verify model assumptions, and we discuss the properties of such techniques.\\
\indent To develop Bayesian model averaged versions of the Sargan test and the 
Cragg and Donald test we rely on the notion of posterior predictive $p$-values. Rubin \shortcite{rubin_1984} argued that posterior predictive $p$-values are useful tools in applied Bayesian statistics to verify model assumptions.  Given a hypothesis, $H$, about the model, the posterior predictive $p$-value given data is calculated as $pr(H|D)$. It quantifies the degree to which the hypothesis is supported by the posterior distribution of the model parameters given the data $D$.\\
\indent In the case of 2SBMA, when the hypothesis relates to the model identification, we consider the following decomposition for model averaged posterior $p$-values:
\begin{equation}\label{eq:mappp}
pr(H|D) = \sum_{i = 1}^{I}\sum_{j = 1}^{J} \pi_i \nu_{i,j} pr(H|M_i,L_j, D).
\end{equation}
The decomposition in Equation~(\ref{eq:mappp}) outlines the motivation for model averaged posterior $p$-values.  Since the hypotheses of primary interest are related to the identification of the 2SBMA estimate, it is natural to ask whether each sub-model which is used to form the 2SBMA estimate is appropriately identified.  Naturally, different specifications yield varying degrees of confidence as to whether the identification requirements are fulfilled and it is appropriate to weight posterior predictive $p$-values for each model by the extent to which each model contributes to the 2SBMA estimate.\\
\subsection{A Bayesian Test of the Over-Identification Restrictions}
\indent By selecting first-stage models based on the integrated likelihood, we are explicitly choosing combinations of instruments based on model fit.  Hall et al. \shortcite{hall_et_1996} note that any procedure that chooses the first stage based on goodness of fit measures may risk including endogenous variables.  Therefore, it is important to develop a diagnostic that helps assess whether the model selection procedure has produced such an over-identification failure.\\
\indent We base our test of over-identification on the Sargan~\shortcite{sargan_1958} test.  Let $\hat{\eta}^{(i,j)}$ be the residuals from the combination of models $M_i$ and $L_j$, and let $p_{i,j}$ be the total number of $X$ and $Z$ included in this combination.  The Sargan $p$-value $S^*$ is calculated as $S^{*} = pr(nR^2_{**} > \chi_{p_X + p_Z - 1}^2)$ where $R^2_{**}$ is the $R^2$ associated with the regression of $\hat{\eta}^{2SLS}$ on all $X$ and $Z$ variables. \\
\indent Just as in the classical Sargan test, we consider the regression of $\hat{\eta}^{(i,j)}$ on the subset of the variables $X$ and $Z$ that belong to either $M_i$ or $L_j$, and determine $R^2_{ij}$, which is the posterior mean $R^2$ associated with this model speficiation.  Letting $S^{(i,j)} = p(\chi_{p_{i,j} - 1}^2 > nR^2_{ij}|X,Y,Z,W)$, where
we define the Bayesian Sargan $p$-value to be
\[
S^{2SBMA} = \sum_{i = 1}^{I} \sum_{j = 1}^{J} \pi_i\nu_{i,j}S^{(i,j)}.
\]
$S^{2SBMA}$ is therefore the average of the Sargan $p$-values derived from the specific models $M_i$ and $L_j$, weighted by their respective posterior probabilities.\\
\indent The benefit of the Bayesian 2SBMA Sargan test is that the parsimony of 2SBMA effectively mitigates the reduction in power that the traditional Sargan test experiences as the dimensions of the $X$ or $Z$ variables grow.  This increase in power can be big, as shown in the simulation study below.  It is important to note that this increase in power derives from the fact that the 2SBMA estimator contains models with greater parsimony than the full specification as opposed to a fundamental difference between Bayesian and frequentist approaches.
\subsection{Bayesian Tests of Under-Identification and Weak Instruments}
\indent While it is crucial to verify that proposed instruments do not violate the conditional independence assumption, it is also important to test that they have explanatory power for the endogenous $W$. When $W$ is univariate, this may be done by considering an $F$ test based on the first stage.  However, when $p_W > 1$, Cragg and Donald \shortcite{cragg_donald_1993} derived an equivalent test for this. Here we derive a Bayesian analog of this test. \\
\indent Consider fixed first and second stage models, $M_i$ and $L_j$ respectively. Let $Z_{ij}$ be the instruments used in this combination, 
namely all those variables in $Z$ used in $M_i$ and those variables $X$ used in $M_i$ but excluded from $L_j$. Let $X_j$ be those $X$ contained in $L_j$, 
and let $V_{ij}$ be the matrix of all $X$ and $Z$ variables included in either $M_i$ or $L_j$.  Define $P_{V_{ij}} \equiv V_{ij}(V_{ij}'V_{ij})^{-1}V_{ij}'$ and $M_{V_{ij}} \equiv I_n - P_{V_{ij}}$ where $I_n$ is the $n\times n$ identity matrix, and similarly define $P_{X_j} \equiv X_j (X_j'X_j)^{-1}X_j'$ and $M_{X_j} = I_n - P_{X_j}$, and finally define $G_{ij} \equiv \hat{\Sigma}^{-1 / 2}_{ij} \Theta_{ij} \hat{\Sigma}^{-1 / 2}_{ij}$ where $\hat{\Sigma}_{ij} = W'M_{V_{ij}}W$ and $\Theta_{ij} = (M_{X_j} W)'M_{X_j} Z_{ij} ((M_{X_j}Z_{ij})'M_{X_j}Z_{ij})^{-1} (M_{X_j}Z_{ij})'M_{X_j}W$. The Cragg and Donald statistic under model $M_i$ and $L_j$ can then be derived as the minimum eigenvalue of $G_{ij}$, $g_{ij} = \min \text{eigen} G_{ij}$.\\
\indent In practice, the statistic $g_{ij}$ is used in two ways.  Asymptotically, under the null hypothesis of under-identification, $n g_{ij} \sim \chi^2_{p_{Z_{ij}} - 1}$, and this reference distribution is used to derive a posterior predictive $p$-value.  Here we propose a Bayesian model-averaged version of this posterior $p$-value by considering
\[
CD = \sum_{i = 1}^{I}\sum_{j = 1}^{J} \pi_i \nu_{i,j} pr(\chi^2_{p_{Z_{ij}} - 1} > n g_{ij}).
\]
\indent A second use of $g_{ij}$ was suggested by Stock and Yogo \shortcite{stock_yogo_2002}, but their test statistic provides only critical values, not the necessary $p$-values that can be averaged when models have different numbers of instruments. The Bayesian approach can also be used to assess an apparent weakness of an instrument. The approach is simple and direct: it simply requires an examination of the instruments' inclusion probabilities.\\

\section{Simulation Study}\label{sec:Sim}
\indent We conduct a simulation study to quantify the estimation properties of 2SBMA and the behavior of the Bayesian tests of the identification restrictions.  

\begin{figure}[tp]
\caption{Finite Sample Bias under 2SBMA, 2SLS and OLS. Distribution of the estimate of the coefficient $\beta_W$ across replications using 2SBMA, 2SLS and OLS, when $\beta_W = 1$.  The average bias of $\hat{\beta}_W$ across 500 replications was $.019$, $.035$ and $.310$ for 2SBMA, 2SLS and OLS respectively.  The average mean squared error for estimating the entire vector $\beta$ was .213, .395 and $.840$ for 2SBMA, 2SLS and OLS respectively.}\label{fig:sim_estimates}
\begin{center}\includegraphics[height = 3in]{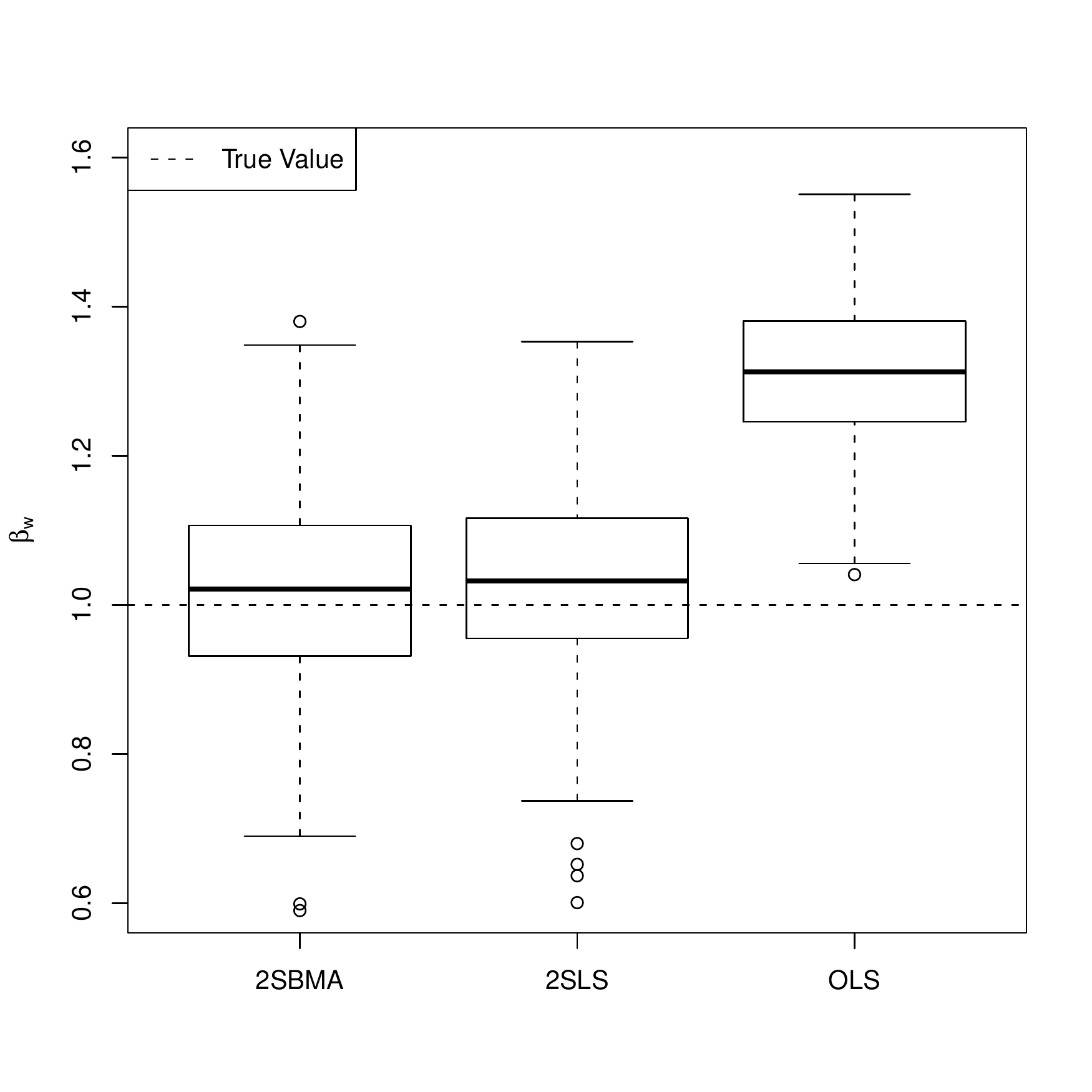}\end{center}
\end{figure}

\begin{figure}[tp]
\caption{Distribution of $p$-values returned by the Bayesian Sargan test and the Sargan test across replications when the IV assumptions hold (left) and when they do not (right[AR2]). In all cases, a nominal level of $5\%$ was used, implying that the assumption of valid instruments would be rejected when the $p$-value was above $.95$, shown by the dotted line in each plot.  In the case of valid instruments, the size of both tests was $0$ (no false rejections).  However, in the case of invalid instruments the power of the Bayesian Sargan test was $50$\% (half the cases were correctly rejected), whereas it was $0.8$\% using the traditional Sargan test (4 cases out of 500 were correctly rejected).}\label{fig:sim_sargan}
\begin{center}\includegraphics[height = 3in]{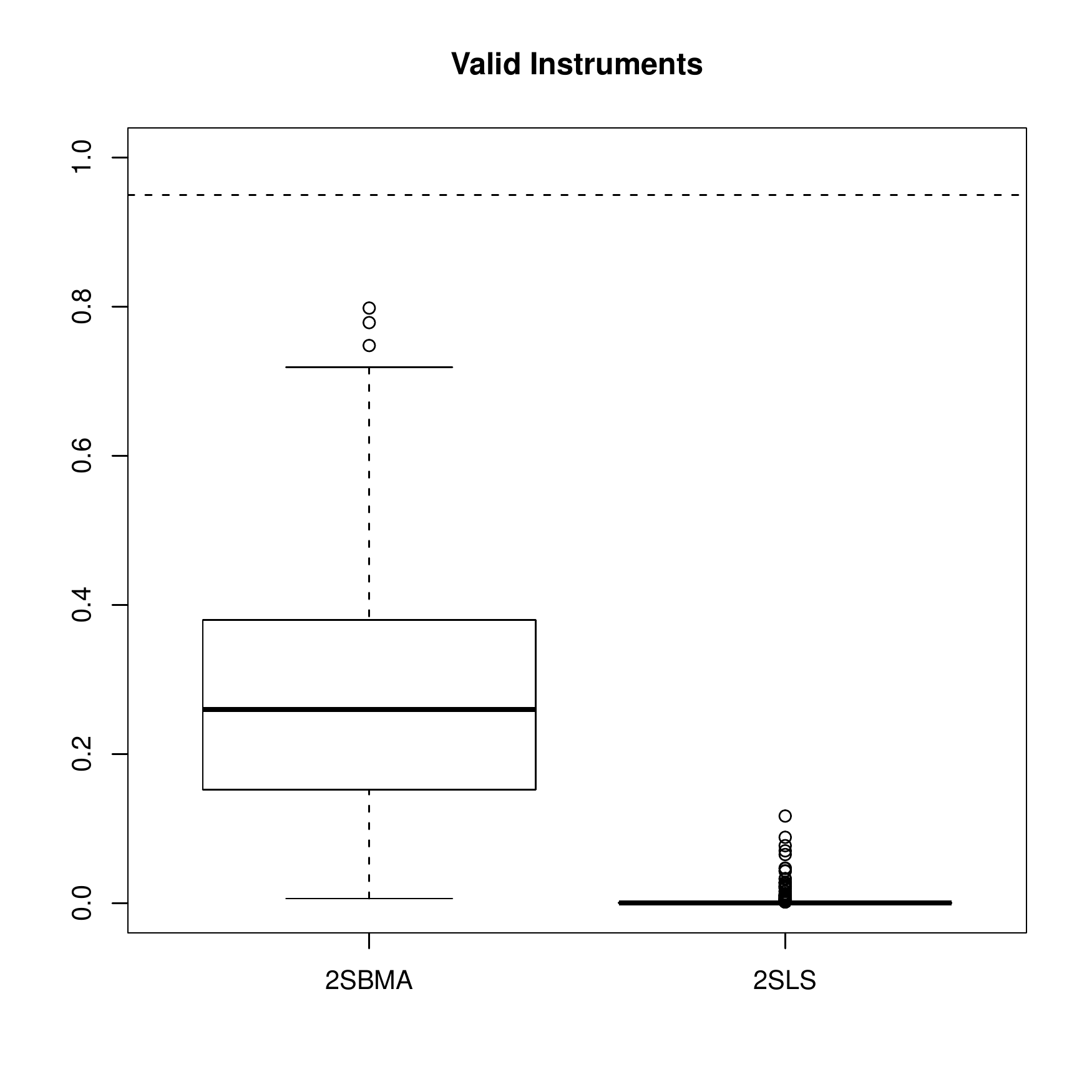}\includegraphics[height = 3in]{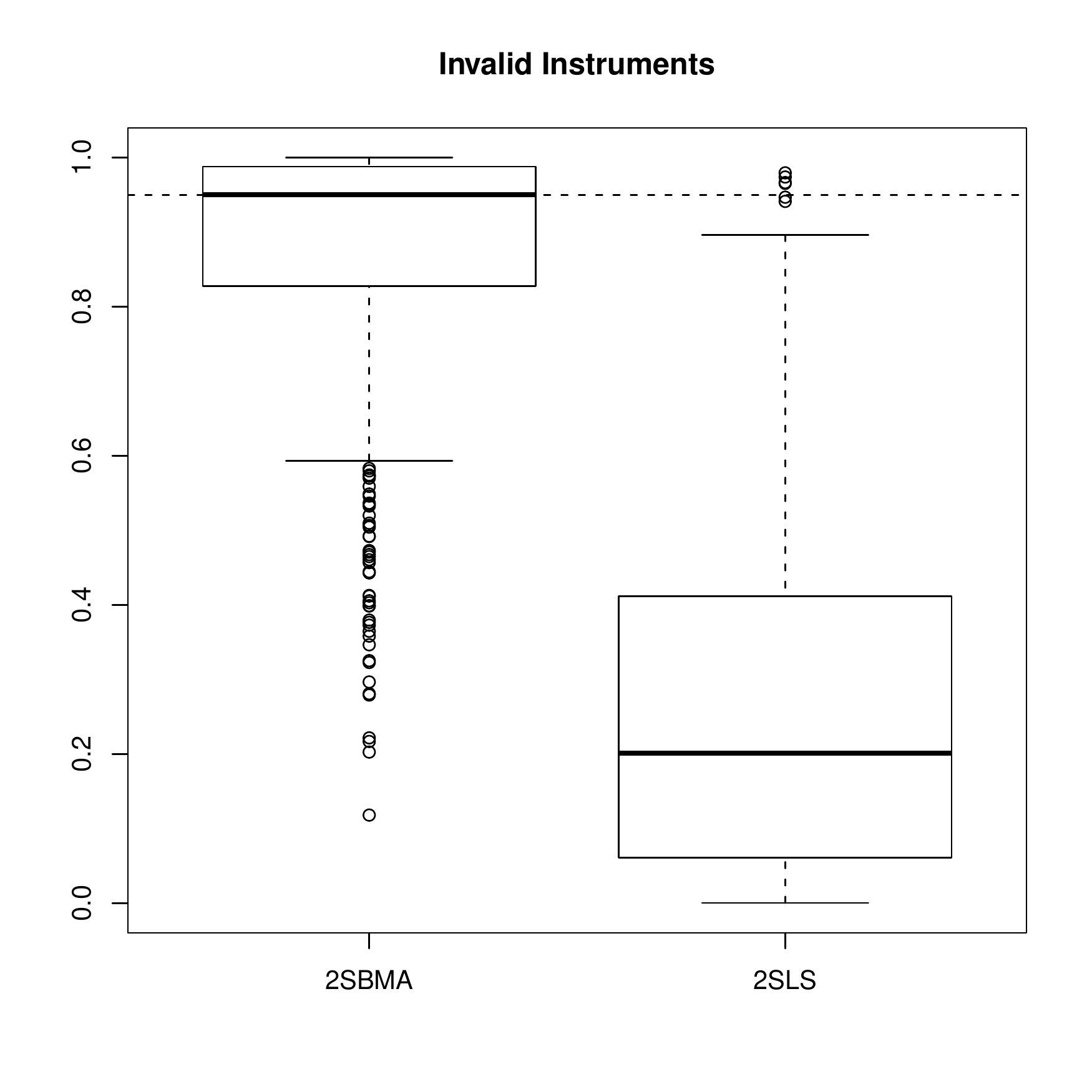}\end{center}
\end{figure}

In the following we consider a framework in which there are ten variables in $Z$, 15 in $X$ and $W$ is univariate.  Our construction of the variables in $X$ and $Z$ is similar to the simulation study in Fernandez et al. \shortcite{fernandez_et_2001}, which was based in turn on Raftery et al. \shortcite{raftery_et_1997}.\\
\indent For constructing $X$ we let $(X_1\mbox{ } \dots\mbox{ }X_{10})$ be an $n \times 10$ matrix of independent draws from a $N(0,1)$ distribution.  We set
\begin{equation}\label{eq:sim_x}
(X_{11}\mbox{ }\dots\mbox{ }X_{15}) = (X_{1}\mbox{ }\dots\mbox{ }X_{5})(.3\mbox{ }.5\mbox{ }.7\mbox{ }.9\mbox{ }1.1)'(1\mbox{ }\dots\mbox{ } 1) + E,
\end{equation}
where $E$ is an $n\times 5$ matrix of draws from the $N(0,1)$ distribution.
Note that Equation (\ref{eq:sim_x}) induces a correlation between the first five regressors and the last five regressors. It takes the form of small to moderate correlations between the first five variables in $X$ and the last five where the the theoretical correlation coefficients increase from 0.153 to 0.561.\\
\indent The matrix $Z$ is sampled in a similar manner.  $(Z_1\mbox{ }\dots\mbox{ }Z_5)$ is an $n \times 5$ matrix of standard normal variates and we then set
\[
(Z_{6}\mbox{ }\dots\mbox{ }Z_{10}) = (Z_{1}\mbox{ }\dots\mbox{ }Z_{5})(.3\mbox{ }.5\mbox{ }.7\mbox{ }.9\mbox{ }1.1)'(1\mbox{ }\dots\mbox{ } 1) + F,
\]
where $F$ is an $n\times 5$ matrix of independent draws from the $N(0,1)$ distribution.  Finally, we construct the model
\begin{align}
Y &= W + 1.8X_1 + 1.5X_2 +  X_{11} -1.5 X_{12} + 2\eta ,\\
W &= 1.1X_1 - .5 X_3 + .75 X_{12} + .75 Z_1 -2 Z_8 + 3\epsilon .
\end{align}
\indent Thus, we consider a situation in which four covariates along with $W$ have explanatory power for $Y$.  
Furthermore, two variables in $Z$ serve as instruments and two of the variables in $X$ have explanatory power on both $Y$ and $W$. 
Finally, one variable in $X$ ($X_3$) would be more properly classified as an instrument, as it has explanatory power on $W$ but not on $Y$.\\
\indent We introduce endogeneity by drawing $\epsilon$ from a $N(0,1)$ distribution and setting $\eta = \epsilon + \xi$, with $\xi$ drawn from a $N(0,1)$ distribution as well. We then consider two scenarios.  The first scenario is one in which the IV model is correctly specified, i.e. the $Z$ covariates have no effect on $Y$. In the second scenario we consider a misspecified model in which $\eta = Z_1 + \epsilon + \xi$, so that the instrument condition fails.  In each scenario we simulate datasets of $100$ observations and consider $500$ replicates.  The simulation study is structured to roughly resemble the data set we will be examining below.\\
\indent Figure \ref{fig:sim_estimates} shows the distribution of the estimate of $\beta_W$ across replications using 2SBMA, 2SLS and OLS.  We see that the OLS estimates are centered about a value of $1.3$.  Indeed, in this case the OLS estimate will asymptotically approach this value.  Both 2SBMA and 2SLS rectify this bias and are more closely centered about the true value of $1$.  
However, the average bias and average mean squared error of 2SBMA are about
45\% lower than those of 2SLS, so 2SBMA performs much better than
OLS or 2SLS. \\
\linespread{1}
\defaultspread
\indent The first panel in Figure~\ref{fig:sim_sargan} shows the distribution 
of the $p$-values returned from the Bayesian Sargan test as well as the 
traditional Sargan test.  We see that the $p$-values from the Bayesian Sargan test are much higher.  
However, these scores are still sufficiently low that the exogeneity assumption is unlikely to be incorrectly rejected.\\
\indent The second panel in Figure~\ref{fig:sim_sargan} shows the resulting Bayesian Sargan and classical Sargan $p$-values for the case of a misspecified exogeneity assumption. 
In the case of valid instruments, the size of both tests was 0. However, in the case of invalid
instruments the power of the Bayesian Sargan test was 50\%, 
whereas it was 0.8\% using the traditional Sargan test,
based on $\alpha = .05$.  
We see that the Bayesian Sargan test is more precise
in discerning the failure of the exogeneity assumption and is far more 
likely to reject the hypothesis that the IV assumptions are valid 
than the classical Sargan test.  As noted above, this result derives
from the fact that the 2SBMA estimates are formed over models with 
considerably greater parsimony than the full specification,
which highlights the benefits of parsimonious models and the
automatic manner in which 2SBMA is able to propose these sparse models.
For the case of valid instruments, both the classical and Bayesian 
Cragg-Donald tests correctly rejected the null hypothesis of no 
identification for all repetitions.\\
\linespread{1}
\defaultspread
\indent The previous figures show that 2SBMA returns appropriate coefficient estimates and yields improved power at detecting assumption violations over traditional methods.  Table~\ref{tab:sim_probs} shows that the technique also uncovers the pattern of interaction in both stages of the estimation.
When the model is correctly specified, Table~\ref{tab:sim_probs} shows the median inclusion probability for each variable across the $500$ replications in both stages as well as the interquartile range of the inclusion probabilities.  We see that in the first stage the two variables in $Z$ as well as the three variables in $X$ are given high inclusion probabilities, while the remaining variables are generally excluded.  This remains true in the second stage as well, where $W$ and the four covariates in $X$ that have explanatory power are given large inclusion probabilities and all others are given negligible inclusion probabilities.\\
\linespread{1}
\begin{table}[p]\caption{Median and interquartile range (IQR) of variable inclusion probabilities across 500 repetitions.  Variables shown in bold are those that are included in either the first or second stage.  This table shows that inclusion probabilities closely match the true structure of the system.}\label{tab:sim_probs}
{\tiny\linespread{1.4}
\begin{center}
\begin{tabular}{l c c c c c}
\hline
 & \multicolumn{2}{c}{First Stage}&&\multicolumn{2}{c}{Second Stage}\\
Variable & $Median$ & IQR && $Median$ & IQR\\
\hline
$W$ &  --  &  --  && \textbf{1} &  \textbf{(1,1)}\\
$X_{1}$ & \textbf{1} &  \textbf{(1,1)} && \textbf{0.761} &  \textbf{(0.573,0.999)}\\
$X_{2}$ & 0.017 &  (0,0.064) && \textbf{0.588} &  \textbf{(0.222,0.962)}\\
$X_{3}$ & \textbf{0.561} &  \textbf{(0.167,0.943)} && 0.065 &  (0.018,0.064)\\
$X_{4}$ & 0.031 &  (0,0.081) && 0.092 &  (0.023,0.096)\\
$X_{5}$ & 0.031 &  (0,0.102) && 0.09 &  (0.024,0.086)\\
$X_{6}$ & 0.019 &  (0,0.082) && 0.091 &  (0.015,0.094)\\
$X_{7}$ & 0.014 &  (0,0.067) && 0.086 &  (0.014,0.083)\\
$X_{8}$ & 0.026 &  (0,0.084) && 0.077 &  (0.018,0.07)\\
$X_{9}$ & 0.018 &  (0,0.077) && 0.077 &  (0.016,0.088)\\
$X_{10}$ & 0.014 &  (0,0.074) && 0.072 &  (0.014,0.065)\\
$X_{11}$ & 0.021 &  (0,0.088) && \textbf{0.473} &  \textbf{(0.1,0.874)}\\
$X_{12}$ & \textbf{1} &  \textbf{(1,1)} && \textbf{0.72} &  \textbf{(0.492,0.987)}\\
$X_{13}$ & 0.016 &  (0,0.076) && 0.094 &  (0.023,0.08)\\
$X_{14}$ & 0.02 &  (0,0.096) && 0.107 &  (0.027,0.109)\\
$X_{15}$ & 0.022 &  (0,0.096) && 0.112 &  (0.027,0.117)\\
$Z_{1}$ & \textbf{0.908} &  \textbf{(0.963,1)} &&  --  &  -- \\
$Z_{2}$ & 0.074 &  (0,0.066) &&  --  &  -- \\
$Z_{3}$ & 0.079 &  (0,0.063) &&  --  &  -- \\
$Z_{4}$ & 0.077 &  (0,0.068) &&  --  &  -- \\
$Z_{5}$ & 0.082 &  (0,0.073) &&  --  &  -- \\
$Z_{6}$ & 0.079 &  (0,0.066) &&  --  &  -- \\
$Z_{7}$ & 0.083 &  (0,0.066) &&  --  &  -- \\
$Z_{8}$ & \textbf{1} &  \textbf{(1,1)} &&  --  &  -- \\
$Z_{9}$ & 0.075 &  (0,0.063) &&  --  &  -- \\
$Z_{10}$ & 0.1 &  (0,0.089) &&  --  &  -- \\
\hline
\end{tabular}
\end{center}
}
\end{table}
\defaultspread
\section{Instrument and Determinant Uncertainty in Development Accounting}
We now apply 2SBMA to one of the most prominent approaches to institutions in the development accounting literature. 
\ffootnote{Our computations use the 
{\tt bicreg} function from the {\tt BMA} R package 
(Raftery et al. \shortcite{Raftery&2005}), since
the Bayesian Information Criteria (BIC) closely approximates the posterior 
model probability under the UIP \cite{kass_wasserman_1995,raftery_1995}.
This allows for BIC approximations of the integrated likelihood in both the first and second stages. For a given value of $\theta^{(i)}$, the quantity $pr(L_j|\theta^{(i)}, M_i, D)$ is given by a regression model with a particular UIP and is therefore well approximated by the BIC.  In addition we can approximate $pr(L_j|M_i,D)$ with $pr(L_j|\hat{\theta}^{(i)}, D)$ and calculate BIC relative to this $\hat{\theta}^{(i)}$.  The conditional independence assumptions of $(M_i, L_j)$ enable us to set $\hat{\Pi}^{(i,j)} = 0$. Thus in our BIC approximation the likelihood $pr(\beta|L_j, \hat{\theta}^{(i)},D)$ is maximized at $\hat{\beta}^{(i,j)}$ defined in Equation~(\ref{eq:const2sls}). Note that BIC can be calculated from the associated $R^2$ of the regression.  By calculating the model probability $\nu_{i,j}$ conditional on the fitted value $V^{(i)}\hat{\theta}^{(i)}$ of $W$, we note that the resulting $R^2$ is equivalent to using the ``Generalized $R^2$" suggested by Pesaran and Smith \shortcite{pesaran_smith_1994} for scoring the second stage model $L_j$.\\}
The literature seeks to explain the variation in the level of per capita income in 1995 with alternative theories of economic development. Rodrik et al. \shortcite{rodrik_et_2004} (RST) juxtapose the most prominent development theories and their associated candidate regressors in one comprehensive approach. They conducted what they explicitly called a ``horse race" among theories that pertain to development determinants (geography, integration/trade and institutions). Endogeneity is a major feature of this literature, since one could also argue that institutions or integration/trade are determined by high levels of per capita income. Hence the development accounting literature has given rise to a large set of alternative theories and their alternative instruments that can be used to identify the effect of institutions and trade on development to resolve the endogeneity issues. This implies that model uncertainty is present in both the development determinant and in the instrument stage. With less than 100 observations, the RST sample is a standard size of datasets in development accounting. \\
\indent RST explored over 25 different robustness specifications with different candidate regressors that are suggested by a comprehensive set of theories that span the literature. Their results are so uniform and decisive across all specifications that their claim to have resolved model uncertainty is well captured by their title: ``Institutions rule: the primacy of institutions over geography and integration in economic development." At best, RST find geography may have weak direct effects while Integration is found to be ``always insignificant, and often enters the income equation with the `wrong' sign." Their results are, however, in stark contrast to previous evidence of Trade/Integration, and Geography effects on development (e.g., Hall and Jones \shortcite{hall_jones_1999}, Sachs \shortcite{sachs_2003}); but these papers were discounted because they did not provide RST's level of robustness analysis.\\
\indent Using RST's own data set, we reexamine their robustness specifications using 2SBMA to address the model and instrument uncertainty that is highlighted so forcefully in their paper. A description of each regressor is provided in Table~\ref{tab:rst_varnames} and the theories that are associated with each regressor are discussed extensively in RST. The 2SBMA first and second stages are reported in Tables~\ref{tab:rst_s1} and~\ref{tab:rst_s2}, respectively.\\
\indent In terms of development determinants, RST assume that Geography is exogenous, so the upper panel in Table~\ref{tab:rst_s1} represents the first stage for the endogenous institutions proxy (Rule of Law) and the lower panel in the same table is the first stage for the endogenous Integration variable.\\
\indent Given the 2SBMA methodology, it would be enough to present only the 2SBMA results that explore the entire model space spanned by RST's development determinants and the associated instruments. This specification is provided in Column 3. We first provide, however, two intermediate stages, where Column 1 represents RST's ``core specification" (RST's Table 2) and Column 2 is the first set of RST's robustness exercises. The three-step approach highlights the sensitivity of the core specification to the introduction of additional covariates that are associated with different theories (in Column 2), as well as sensitivity to different variables associated with the different theories (in Column 3).\\
\indent Column 1 in Table~\ref{tab:rst_s2} provides the second stage of RST's preferred core specification (RST's Table 2). Both RST and 2SBMA find that only Rule of Law shows an effect and the conditional posterior mean is nearly identical to RST's 2SLS estimate. In this specification, the 2SBMA result confirms RST's central finding that ``the preferred specification accounts for about half of the variance in incomes across the sample, with institutional quality (instrumented by settler mortality) doing most of the work." The generalized $R^2$ for the best 2SBMA model is 0.53 versus 0.55 in RST's 2SLS approach.\\
\indent Column 1 in Table~\ref{tab:rst_s1} reports the two 2SBMA first stages for RST's core specification. They broadly confirm the 2SLS results in RST although 2SBMA suggests slightly more parsimonious models. 2SBMA suggests three strong instruments for Rule of Law (Settler Mortality, Latitude, and the Fraction Speaking English), while RST found significant coefficients for all five instruments across their various 2SLS exercises. This generates a slightly higher $R^2$ for RST's preferred 2SLS specification (0.55) as compared to the best model in 2SBMA (0.49). For Integration, the 2SBMA first stage suggests only two strong instruments (Implied Trade Shares and Settler Mortality) while 2SLS produces a statistically significant coefficient for an additional instrument (Fraction Speaking English). Nevertheless the $R^2$ of the 2SBMA best model and of the 2SLS first stage are identical (0.58).\\
\indent RST find that core specifications with more than one instrument fail to pass the Sargan test. This finding is confirmed by the Bayesian Sargan test in Column 1 of Table~\ref{tab:rst_s2}, which presents a similar $p$-value to that found by RST. One interpretation is that the Sargan test undermines alternative determinant and instrument strategies as suggested by RST.  One could also argue, however, that RST's specification does not contain the appropriate set of instruments. We examine this issue further below but note that already at this stage, under-identification (as measured by the Bayesian Cragg-Donald $p$-value) is easily rejected by 2SBMA (not reported in RST). Weak instruments are not of concern in this data set.\\
\indent The 2SLS results in RST's core specification and the 2SBMA results in Column 1 are nearly identical because the core specification includes minimal model uncertainty at the determinant level and only a fraction of the standard instruments suggested by the development literature. Column 2 adds regressors suggested by alternative theories that pertain to Legal Origins and Religion, as well as regional dummies, while Column 3 represents the most comprehensive set of regressors that adds standard covariates related to alternative Geography theories (most notably Temperature, Malaria) as well as alternative Integration measures (such as Sea Access). As we allow for additional theories and the associated regressors, the 2SBMA results start to diverge from the results in RST's individual 2SLS robustness regressions. In other words, the disparities across results become more pronounced as model uncertainty increases. \\
\indent The 2SBMA results that use the most comprehensive set of instruments and development determinants (Column 3 in Tables~\ref{tab:rst_s1} and~\ref{tab:rst_s2}), cast doubt on the primacy of institutions result. Instead 2SBMA finds that the ``horse race'' ends in a statistical three-way tie. Geography (as measured by Tropics), Institutions and Integration are shown to be highly effective development determinants. This result is particularly surprising in light of the fact that Geography is only occasionally weakly significant in RST, and Integration is never significant and often of the wrong sign. In 2SBMA all three effects are strong and estimated with the correct sign. Once model uncertainty is comprehensively addressed at both the development determinant, as well as the instrument level, the results thus support the contentions of Sachs \shortcite{sachs_2003} and Alcal\'{a} and Ciccone \shortcite{alcala_ciccone_2004}, who report strong effects of Geography and Integration. \\
\indent The divergence of 2SLS and 2SBMA results can already be observed in the first stages.\ffootnote{RST report neither first stages nor tests of instrument restrictions beyond the core specification.} Most importantly, the Implied Trade Share no longer receives support as a strong instrument for Integration. It is most strongly instrumented by EuroFrac in combination with the covariates PopGrowth, Oil, SeaAccess, Malaria94, EuroFrac, Tropics, Latitude, FrostArea, and PolicyOpenness. In contrast to the findings of RST, religion variables also play an important part in the first stage regression.  In particular, Catholic has nearly a 90 percent inclusion probability in the first stage for Rule of Law and above 50 percent in the first stage for Integration.  Similarly, the power of Settler Mortality as an instrument for Institutions is dominated by regressors such as EuroFrac and Temperature variables in both first stages. \\ 
\indent Note that the increase in the model space of development determinants and instruments going from the specification in Column 1 to Column 3 dramatically improves the fit of the 2SBMA first stage. For Column 3, the best models in both 2SBMA first stages report an $R^2$ that is at least 40 percent greater than those found in the core specification.\\
\indent A similar improvement in model fit can be observed in the second stage when considering the generalized $R^2$ \cite{pesaran_smith_1994} of the best model returned by 2SBMA. In fact, none of the top 100 models' generalized $R^2$ falls below .82, which greatly exceeds any model presented by RST (whose highest generalized $R^2$ is .73). 2SBMA has therefore uncovered combinations of instruments and development determinants that fit the data substantially better. This is then the source of the difference in the 2SLS and 2SBMA results in both the first and second stages. The Bayesian Sargan and Bayesian Cragg and Donald tests clearly show, respectively, that over-identification is easily rejected with the improved set of instruments and that under-identification remains of little concern.\\
\linespread{1}
\begin{table}\caption{Variable Descriptions from RST dataset.}\label{tab:rst_varnames}
{\tiny
\begin{tabular}{l l}
\hline
Variable Name & Description\\
\hline
Area & Land area (thousands sq. mt.) \\
Catholic & Dummy variable taking value 1 if the country’s population is predominantly catholic\\
EastAsia & Dummy variable taking value 1 if a country belongs to South-East Asia, 0 otherwise\\
EngFrac & Fraction of the population speaking English. \\
FR\_tradeShares & Natural logarithm of predicted trade shares computed from a bilateral trade equation with “pure geography” variables. \\
FrostArea & Proportion of land with >5 frost-days per month in winter. \\
FrostDays & Average number of frost-days per month in winter. \\
Integration & Natural logarithm of openness.  Openness is given by the ratio of (nominal) imports plus exports to GDP (in nominal US dollars).\\
LatinAmerica & Dummy variable taking value 1 if a country belongs to Latin America or the Caribbean, 0 otherwise\\
Latitude & Distance from Equator of capital city measured as abs(Latitude)/90\\
LegalOrigFr & variable taking a value of 1 if a country has a legal system deriving from that in France\\
LegalOrigSocialist & variable taking a value of 1 if a country has a socialist legal system\\
Malaria94 & Malaria index, year 1994. \\
MeanTemp & Average temperature (Celsius). \\
Muslim & Dummy variable taking value 1 if the country’s population is predominantly muslim\\
Oil & variable taking value 1 for a country being major oil exporter, 0 otherwise.  \\
PolicyOpenness & Dummy variable that indicates if a country has sufficiently market oriented policies\\
PopGrowth & population growth\\
Protestant & variable taking value 1 if the country’s population is predominantly protestant\\
RuleofLaw & Rule of Law index. Refers to 2001 and approximates for 1990’s institutions \\
SeaAccess & Dummy variable taking value 1 for countries without access to the sea, 0 otherwise.  \\
SettlerMortality & Natural logarithm of estimated European settlers’ mortality rate\\
SubSaharaAfrica & taking value 1 if a country belongs to Sub-Saharan Africa, 0 otherwise\\
Tropics & Percentage of tropical land area.\\
\hline
\end{tabular}
}
\end{table}

\begin{table}[p]\caption{First Stage Results for RST Example}\label{tab:rst_s1}
{\tiny
\begin{tabular}{|l |c c c| c c c |c c c|}
\hline
& \multicolumn{3}{|c|}{I} & \multicolumn{3}{|c|}{II} & \multicolumn{3}{|c|}{III}\\
& \multicolumn{3}{|c|}{RST Table 2} & \multicolumn{3}{|c|}{RST Table 2, 4} & \multicolumn{3}{|c|}{RST Table 2, 4, 5, 6}\\
&\multicolumn{3}{|c|}{Core Specification} & \multicolumn{3}{|c|}{I + LegalOrig, Relig, Region} & \multicolumn{3}{|c|}{II + Alt. Integr./Geo Measures}\\
\hline
&\multicolumn{9}{|c|}{\emph{Stage 1, Dependent Variable: Rule of Law}}\\
& $p\neq 0$ & Mean & Sd & $p\neq 0$ & Mean & Sd & $p\neq 0$ & Mean & Sd\\
\hline
SettlerMortality & 92.7 & -0.21189 & 0.07113 & 25.5 & -0.10696 & 0.07218 & 17.1 & -0.02528 & 0.06317 \\
EuroFrac & 16.5 & 0.23 & 0.23093 & 99.9 & 1.66071 & 0.31244 & 100 & 1.03 & 0.2994 \\
Catholic &  &  &  & 14.9 & -0.00478 & 0.00381 & 89.9 & -0.01405 & 0.00576 \\
MeanTemp &  &  &  &  &  &  & 86.8 & -0.05535 & 0.02794\\ 
PopGrowth &  &  &  &  &  &  & 72.5 & -0.1011 & 0.08351 \\ 
SubSaharaAfrica &  &  &  & 10.7 & -0.34126 & 0.30796 & 55.5 & -0.2287 & 0.2442 \\
Muslim &  &  &  & 13 & -0.00434 & 0.00393 & 40.2 & -0.00204 & 0.003 \\
Latitude & 89.7 & 0.02254 & 0.00792 & 99.1 & 0.02919 & 0.00732 & 20.5 & 0.00411 & 0.00938 \\
LatinAmerica &  &  &  & 99.9 & -1.00161 & 0.27639 & 14.9 & -0.1277 & 0.3428 \\
Area &  &  &  &  &  &  & 12.8 & .02582 & .07443 \\
Oil &  &  &  &  &  &  & 8.8 & -0.03573 & 0.1377 \\
FR\_Trade Shares & 38.5 & 0.17551 & 0.09726 & 99 & 0.28821 & 0.08545 & 8 & -0.01918 & 0.08153 \\
Tropics &  &  &  &  &  &  & 7.9 & -0.02392 & 0.1062 \\
EngFrac & 98.6 & 1.07778 & 0.28172 & 12.2 & 0.37306 & 0.35301 & 7.5 & 0.04617 & 0.201 \\
FrostArea &  &  &  &  &  &  & 6.3 & 0.04695 & 0.2127 \\
Protestant &  &  &  & 10.7 & -0.00702 & 0.00619 & 3.8 & 0.00026 & 0.00159 \\
FrostDays &  &  &  &  &  &  & 1.9 & 0.00043 & 0.00725 \\
LegalOrigFr &  &  &  & 46.7 & -0.30591 & 0.15008 & 1.8 & -0.00387 & 0.03538 \\
SeaAccess &  &  &  &  &  &  & 1.4 & 0.00211 & 0.02705 \\
PolicyOpenness &  &  &  &  &  &  & 1.1 & 0.00268 & 0.03683 \\
EastAsia &  &  &  & 91.4 & 0.72988 & 0.25051 & 0 & 0 & 0 \\
Malaria94 &  &  &  &  &  &  & 0 & 0 & 0 \\
LegalOrigSocialist &  &  &  & 63.9 & -0.77728 & 0.36386 & na & na & na\\ 
\hline
BIC best model & -41.53 &  &  & -53.81 &  &  & -51.42 &  &  \\
R2 best model & 0.49 &  &  & 0.66 &  &  & 0.75 &  &  \\
\hline
&\multicolumn{9}{|c|}{\emph{Stage 1, Dependent Variable: Integration}}\\
& $p\neq 0$ & Mean & Sd & $p\neq 0$ & Mean & Sd & $p\neq 0$ & Mean & Sd\\
\hline
FR\_Trade Shares & 100 & 0.5985 & 0.0612 & 100 & 0.5769 & 0.0502 & 0.7 & -0.086 & 0.1074\\
LegalOrigSocialist &  &  &  & 20.1 & -0.2561 & 0.2001 & na & na & na \\
PopGrowth &  &  &  &  &  &  & 100 & -0.2735 & 0.0285 \\
SeaAccess &  &  &  &  &  &  & 94.8 & -0.3023 & 0.106 \\
Oil &  &  &  &  &  &  & 94.4 & 0.3445 & 0.1284 \\
Malaria94 &  &  &  &  &  &  & 91.8 & -0.4383 & 0.1399 \\
EuroFrac & 14.4 & -0.1053 & 0.1389 & 6.1 & 0.0563 & 0.1329 & 81.2 & -0.5145 & 0.1826 \\
Tropics &  &  &  &  &  &  & 73.1 & 0.4392 & 0.1921 \\
Latitude & 23.5 & -0.0065 & 0.005 & 4.9 & 0.0003 & 0.0039 & 72.7 & -0.0164 & 0.007 \\
FrostArea &  &  &  &  &  &  & 65.3 & 0.497 & 0.2019 \\
PolicyOpenness &  &  &  &  &  &  & 59.1 & 0.3468 & 0.1391 \\
Catholic &  &  &  & 7.3 & 0.001 & 0.0014 & 52.5 & -0.0036 & 0.0018 \\
SettlerMortality & 84.9 & -0.1111 & 0.0408 & 9 & -0.0349 & 0.0371 & 50.2 & -0.1077 & 0.0579 \\
EastAsia &  &  &  & 100 & 0.8236 & 0.139 & 28.2 & 0.2917 & 0.1663 \\
EngFrac & 23.2 & 0.246 & 0.1865 & 83.6 & 0.382 & 0.1431 & 24.7 & -0.6486 & 0.3051 \\
FrostDays &  &  &  &  &  &  & 17.1 & 0.0217 & 0.0125 \\
LatinAmerica &  &  &  & 6.3 & -0.0448 & 0.1147 & 16.6 & -0.4149 & 0.1899 \\
MeanTemp &  &  &  &  &  &  & 13.5 & -0.0225 & 0.0118 \\
SubSaharaAfrica &  &  &  & 5.3 & -0.033 & 0.0925 & 4 & -0.1922 & 0.1627 \\
LegalOrigFr &  &  &  & 6 & 0.0464 & 0.1039 & 3.8 & -0.095 & 0.0902 \\
Protestant &  &  &  & 11.9 & 0.0041 & 0.0035 & 0.7 & 0.0024 & 0.0029 \\ 
Muslim &  &  &  & 5.3 & -0.0005 & 0.0012 & 0.2 & -0.0007 & 0.0015 \\
Area &  &  &  &  &  &  & 0.2 & 0 & 0 \\
\hline
BIC best model & -61.22 &  &  & -84.37 &  &  & -54.57 &  &  \\
R2 best model & 0.58 &  &  & 0.71 &  &  & 0.81 &  &  \\
\hline
\end{tabular}
}
\end{table}
\newpage
\begin{table}\caption{Second Stage Results for RST Example}\label{tab:rst_s2}
{\tiny
\begin{tabular}{|l |c c c| c c c |c c c|}
\hline
& \multicolumn{3}{|c|}{I} & \multicolumn{3}{|c|}{II} & \multicolumn{3}{|c|}{III}\\
& \multicolumn{3}{|c|}{RST Table 2} & \multicolumn{3}{|c|}{RST Table 2, 4} & \multicolumn{3}{|c|}{RST Table 2, 4, 5, 6}\\
&\multicolumn{3}{|c|}{Core Specification} & \multicolumn{3}{|c|}{I + LegalOrig, Relig, Region} & \multicolumn{3}{|c|}{II + Alt. Integr./Geo Measures}\\
\hline
&\multicolumn{9}{|c|}{\emph{Stage 2, Dependent Variable: log GDP per capita in 1995}}\\
& $p\neq 0$ & Mean & Sd & $p\neq 0$ & Mean & Sd & $p\neq 0$ & Mean & Sd\\
\hline
Rule of Law & 100 & 1.2775 & 0.1772 & 100 & 0.9485 & 0.1323 & 96.4 & 0.7979 & 0.3155  \\
Integration & 20 & 0.1119 & 0.2578 & 7.4 & 0.0697 & 0.1451 & 84.7 & 0.9275 & 0.3803  \\
Tropics &  &  &  &  &  &  & 69 & -0.7828 & 0.37  \\
Area &  &  &  &  &  &  & 57.1 & .164 & .171  \\
SubSaharaAfrica &  &  &  & 97 & -0.7487 & 0.1998 & 50.7 & -0.5319 & 0.3077  \\
Catholic &  &  &  & 36.2 & 0.0043 & 0.0028 & 50.6 & 0.01 & 0.0072  \\
PolicyOpenness &  &  &  &  &  &  & 49.4 & 0.6857 & 0.368  \\
PopGrowth &  &  &  &  &  &  & 46.7 & 0.2099 & 0.1473  \\
Muslim &  &  &  & 50.3 & -0.0044 & 0.0025 & 43.8 & -0.0043 & 0.0035  \\
LatinAmerica &  &  &  & 10.1 & 0.0984 & 0.2858 & 36.1 & 0.6529 & 0.3652  \\
LegalOrigFr &  &  &  & 29.5 & 0.2083 & 0.2065 & 34.6 & 0.29 & 0.1682  \\
FrostArea &  &  &  &  &  &  & 33.3 & 1.2204 & 0.8814  \\
FrostDay &  &  &  &  &  &  & 31.3 & -0.0621 & 0.0383  \\
MeanTemp &  &  &  &  &  &  & 22.2 & 0.0323 & 0.0433  \\
EastAsia &  &  &  & 22.8 & 0.3345 & 0.3127 & 19.5 & 0.532 & 0.3898  \\
Latitude & 18.3 & -0.0019 & 0.0143 & 10.8 & -0.0058 & 0.0099 & 18.6 & -0.0168 & 0.0162  \\
Oil &  &  &  &  &  &  & 18 & 0.323 & 0.2919  \\
Malaria94 &  &  &  &  &  &  & 7.3 & -0.243 & 0.4787  \\
SeaAccess &  &  &  &  &  &  & 5.6 & -0.0698 & 0.3142  \\
Protestant &  &  &  & 8 & -0.0027 & 0.006 & 1.9 & -0.0016 & 0.0069  \\
LegalOrigSocialist &  &  &  & 41 & -0.6144 & 0.4917 & na & na & na  \\
\hline
BIC best model & -57.34 &  &  & -92.34 &  &  & -77.12 &  &   \\
Generalized R2 best model & 0.53 &  &  & 75.10 &  &  & 85.70 &  &   \\
Bayes/Sargan $p$-value & 0.0308 &  &  & 0.7581 &  &  & 0.8438 &  &   \\
Bayes/Cragg-Donald $p$-value & 0.0000 &  &  & 0.0000 &  &  & 0.0097 &  &   \\
\hline
\end{tabular}
}
\end{table}
\defaultspread
\section{Conclusion}
We have developed a methodology to address model uncertainty in the presence of endogeneity and explored its properties as a valid IV estimator. The method is based on Bayesian Model Averaging (BMA), which has already been extensively used in economic applications, particularly in modeling economic growth and development. 2SBMA is shown to be a consistent methodology that merges the BMA and 2SLS procedures in a natural and straightforward manner.\\
\indent In order to extend the 2SLS paradigm to a Bayesian context, we introduced a Two-Stage UIP, which built on the development of the Bayesian two-stage prior of Kleibergen and Zivot \shortcite{kleibergen_zivot_2003}.  This allowed us to leverage existing widely-used software for regression variable uncertainty \cite{Raftery&2005}.\\
\indent A key assumption in our development was that the errors were normally distributed.  Such an assumption is not necessary in order for standard 2SLS to function properly as a consistent estimator, but it was important in specifying our two-stage UIP and is similar to most Bayesian developments of the endogenous variable model \cite{dreze_1976,kleibergen_zivot_2003}.  Relaxing the normality assumption therefore constitutes an important further step in the development of this methodology.  One potential avenue for performing this relaxation is to consider extending the Dirichlet process approach employed by Conley et al. \shortcite{conley_et_2008}.\\
\indent Instrumental variable estimation of any kind requires a number of assumptions that relate to the identification of the implied structural model.  To examine the validity of these assumptions in the context of 2SBMA, we propose Bayesian Sargan and Bayesian Cragg and Donald tests of the over- and under-identification restrictions. These tests are based on model averaged posterior $p$-values within the 2SBMA framework.  We show that the power of these tests is substantially greater 
than those of standard 2SLS tests using the full specification. In addition, within the context of 2SBMA these tests are less affected by an increase in the number of potential instruments. We have focused on deriving these particular tests in part because they are well-known to the majority of applied practitioners.  Furthermore, we have shown that our implementation of these tests, in particular the Bayesian Sargan test, performs in a satisfactory manner.  Alternative frameworks of verifying model assumptions based more directly on the comparison of Bayes Factors, or tests that deal with more complicated sampling assumptions, for instance heteroskedasticity could have also been proposed.  It will be an interesting and important development to see what additional insight such tests could provide.\\
\indent We apply 2SBMA to a prominent development accounting approach (Rodrik et al. \shortcite{rodrik_et_2004}), which was itself motivated by the vast model uncertainty associated with alternative theories of development and alternative instruments to control for potential endogeneity. Instead of resolving the model uncertainty in a horse race of alternative regressions, we use the formal 2SBMA approach. We find not only support for institutions, but also substantial support for geographic and trade factors, once model uncertainty in the presence of endogeneity is addressed. The latter two effects had been relegated to second order effects by RST.\\
\indent The purpose of our paper is to introduce the statistical foundations of 2SBMA methodology and provide applications that highlight the importance of instrument and covariate uncertainty in economics. In the context of the RST example, alternative approaches to development accounting include Mauro \shortcite{mauro_1995}, who first suggested ethnolinguistic fragmentation as a fundamental determinant of corruption, and Hall and Jones \shortcite{hall_jones_1999}, who introduced Latitude and Language indicators as instruments to measure Western influence. Acemoglu et al. \shortcite{acemoglu_et_2001b,acemoglu_et_2001} suggested population density in 1500 and the type of colonial origin (indicated by settler mortality) as effective instruments, respectively. La Porta et al. \shortcite{laporta_et_2004} presented yet another "horse race" of theories, in their case juxtaposing judicial independence and constitutional review. In RST the "horse race" is between three possible determinants: Institutions, Integration, and Geography. Geography-based theories of fundamental development determinants have previously been proposed by Bloom and Sachs \shortcite{bloom_sachs_1998}, Easterly and Levine \shortcite{easterly_levine_2003}, and Sachs \shortcite{sachs_2003}. \\
\indent Several modifications of the BMA paradigm have been 
proposed in the development determinant literature.
Brock et al. \shortcite{brock_et_2003} and Durlauf et al. \shortcite{durlauf_et_2008} discuss priors on the model space that account for the fact that many variables may be collected to proxy one particular theory, while fewer may be available to proxy an alternative theory.  Ley and Steel \shortcite{ley_steel_2007} and Doppelhofer and Weeks \shortcite{doppelhofer_weeks_2009} develop metrics to quantify the degree to which development determinants act ``jointly'' to affect growth.  Determining how these extensions of the BMA paradigm may be taken into account in the 2SBMA framework would help extend the application of 2SBMA to the particular problem of testing growth theory robustness.\\
\indent The 2SBMA method allows researchers to incorporate concepts of model uncertainty and model averaging into the assessment of a diverse range of economic behavior where observations are subject to endogeneity.  However, the current framework does not directly handle such concepts as panel data, mixed effects, random coefficient models, and heteroskedasticity.  Future research into these areas will improve the applicability of the BMA framework to economic analysis, in growth economics and beyond.

\clearpage

\bibliographystyle{jasa}
\bibliography{ivbma-20110701aer.bib}
\newpage
\renewcommand{\theequation}{A-\arabic{equation}} 
\setcounter{equation}{0}  
\section*{Appendix}
\indent \emph{Proof of Theorem~\ref{thm:MLE}}: In what follows, for a matrix $A$, let $\bfP_{A} = A(A'A)^{-1}A'$ and $\bfM_{A} = I - \bfP_{A}$.  The likelihood equations follow directly from the marginal distribution of $\epsilon$ and the conditional distribution of $\xi$ given $\epsilon$.  Furthermore, the fact that $\hat{\theta}$ maximizes Equation~\ref{eq:marginalW} follows directly from standard results for the MLE of a regression equation.\\
\indent Now, consider $\beta$.  Note that if $\hat{U}$ is in the column space of $V$, which occurs in this case, then $\hat{U} = \bfP_{V}U$. By moving to log likelihoods, and differentiating with respect to the vector $\beta$ we see that
\begin{equation}\label{eq:MLEfirst}
\beta = (\hat{U}'\hat{U})^{-1}\hat{U}'Y + \phi(\hat{U}'\hat{U})^{-1}\hat{U}'\hat{\epsilon}
\end{equation}
but note that
\[
\hat{U}'\hat{\epsilon} = U'\bfP_{V}\bfM_{V}V = 0
\]
Since $\bfP_{V}\bfM_{V} = 0$, thus Equation~\ref{eq:MLEfirst} becomes $\beta = \hat{\beta}^{2SLS}$.$\qed$\\

\indent \emph{Proof of Theorem~\ref{thm:MLEConstr}}: This follows directly from the proof of Theorem~\ref{thm:MLE}.  However, we note that it is possible that $\hat{U}^{(j)}$ may not be written as $\bfP_{V^{(i)}}U^{(j)}$ for arbitrary $M_i$ and $L_j$, which causes the additional $\phi^{(i,j)}\Pi^{(i,j)}$ to be present.$\qed$\\

\indent \emph{Proof of Theorem~\ref{thm:2SBMA_var}}:  Note that using the standard BMA results, the variance of $\hat{\beta}^{2SBMA}$ can be written as
\begin{equation}\label{eq:2SBMA_var}
\sigma^2_{2SBMA}(\beta) = \sum_{i = 1}^{I}\sum_{j = 1}^{J}\pi_i\nu_{i,j}Var(\hat{\beta}^{(i,j)}) + \sum_{i = 1}^{I}\sum_{j = 1}^{J} \pi_i \nu_{i,j}(\hat{\beta}^{(i,j)} - \hat{\beta}^{2SBMA})^2.
\end{equation}
Rewriting this we have,
\begin{align}
\sigma^2_{2SBMA}(\beta) &= \sum_{i = 1}^{I} \pi_i \left\{\sum_{j = 1}^{J}\nu_{i,j}\left[Var(\hat{\beta}^{(i,j)}) + (\hat{\beta}^{(i,j)} - \hat{\beta}^{2SBMA})^2\right]\right\},\\
&= \sum_{i = 1}^{I}\pi_i \left\{\sum_{j = 1}^{J}\nu_{i,j} \left[Var(\hat{\beta}^{(i,j)}) + (\hat{\beta}^{(i,j)} - \hat{\beta}^{(i,\cdot)} + \hat{\beta}^{(i,\cdot)} - \hat{\beta}^{2SBMA})^2\right]\right\},\\
&= \sum_{i = 1}^{I}\pi_i \left\{\sum_{j = 1}^{J}\nu_{i,j}\left[Var(\hat{\beta}^{(i,j)}) + (\hat{\beta}^{(i,j)} - \hat{\beta}^{(i,\cdot)})^2 + (\hat{\beta}^{(i,\cdot)} - \hat{\beta}^{2SBMA})^2\right]\right\},
\end{align}
which results since,
\[
\sum_{j = 1}^{J}\nu_{i,j}(\hat{\beta}^{(i,j)} - \hat{\beta}^{(i,\cdot)})(\hat{\beta}^{(i,\cdot)} - \hat{\beta}^{2SBMA}) = 0.
\]
Reordering the terms we then obtain
\[
\sigma^2_{2BMA}(\beta) = \sum_{i = 1}^{I}\pi_i Var(\beta|M_i) + \sum_{i = 1}^{I} \pi_i (\hat{\beta}^{(i,\cdot)} - \hat{\beta}^{2SBMA})^2,
\]
as desired.$\qed$\\

\indent \emph{Proof of Theorem~\ref{thm:consistency}}: For convenience, suppose that $M_1 \in \mcl{M}$ is the true model for the first stage.  By true model, we do not mean the underlying model that gave rise to $W$, but the correct subset of $V$ for which the associated elements in $\theta$ are not zero. Then,
\[
\pi_1 \to_p 1 \text{ and }\pi_j \to_p 0, j\neq 1\text{ as }n\to\infty.
\]
by the consistency of the model selection procedure. Furthermore, suppose that $L_1 \in\mcl{L}$ is the true second stage model.  Then,
\[
\nu_{1,1} \to_p 1 \text{ and }\nu_{1,j} \to_p 0, j\neq 1\text{ as }n\to\infty.
\]
Therefore,
\[
\hat{\beta}^{2SBMA} \to_p \hat{\beta}^{(1,1)}.
\]
Finally consider $\hat{\beta}^{2SLS}$.  We know that $\hat{\beta}^{2SLS}\to_p\beta$ by the consistency of the technique.  Furthermore, since the first and second stage estimates of 2SLS are individually consistent we have $\hat{\beta}^{2SLS} \to_p \hat{\beta}^{(1,1)}$ provided $M_1$ and $L_1$ are the true first and second stage models. Thus, $\hat{\beta}^{(1,1)} \to_p \beta$, showing the technique is consistent.$\qed$\\

\end{document}